%% file: main.tex
\documentclass[conference,compsoc]{IEEEtran}
\usepackage{tikz}
\usepackage{amsmath}

\input{pages/config.tex}

\begin{document}

\date{}

\title{Provable Execution in Real-Time Embedded Systems}

\author{
{\rm Antonio Joia Neto}\\
Rochester Institute of Technology\\aj4775@rit.edu
\and
{\rm Norrathep Rattanavipanon }\\
Prince of Songkla University\\norrathep.r@phuket.psu.ac.th
\and
{\rm Ivan De Oliveira Nunes}\\
Rochester Institute of Technology\\ivanoliv@mail.rit.edu
} 

\thispagestyle{plain}
\pagestyle{plain}
\maketitle
\pagenumbering{gobble}
\begin{abstract}
\input{pages/0_Abstract.tex}

\end{abstract}

\input{pages/999_Official_Main}
\bibliographystyle{IEEEtran}
\bibliography{main.bib}

\input{pages/15_Appendix}



\section{META-REVIEW}

The following meta-review was prepared by the program committee for the 2025
IEEE Symposium on Security and Privacy (S\&P) as part of the review process as
detailed in the call for papers.

\subsection{Summary}
This paper develops a provable execution (PoX) system that aims to enable interrupt, a necessary condition for real-time availability, which was not never supported in prior efforts. 

\subsection{Scientific Contributions}
\begin{itemize}
\item Provides a Valuable Step Forward in an Established Field
\item Creates a New Tool to Enable Future Science
\item Addresses a Long-Known Issue
\end{itemize}

\subsection{Reasons for Acceptance}
\begin{enumerate}
\item Real-time availability is a long standing challenge and this paper provides a value step in harmonizing real-time availability with the existing line of work on proof of execution
\item The solution is implemented and evaluated on actual hardware.
\end{enumerate}


\end{document}

%% file: pages/config.tex
\usepackage{nccmath}
\usepackage{soul}
\usepackage{array}
\usepackage{paralist}
\usepackage{tabularx}
\usepackage{etoolbox,xspace}
\usepackage{mdframed}
\usepackage{lipsum}
\usepackage[normalem]{ulem}
\usepackage{float}
\usepackage{amssymb}
\definecolor{littlegreen}{RGB}{193,236,193}
\definecolor{littlegrey}{RGB}{128,128,128}
\definecolor{altgreen}{RGB}{0,150,0}
\usepackage{booktabs}
\usepackage{graphicx}
\usepackage{multirow}
\usepackage{listings}
\usepackage{tcolorbox}

\usetikzlibrary{arrows}
\usetikzlibrary{shapes}

\renewcommand*\circ[1]{\tikz[baseline=(char.base)]{\node[scale=.8,shape=circle,draw,inner sep=1pt,fill=black,text=white] (char) {#1};}}

\usepackage{color,colortbl}
\usepackage{enumitem}

\newlist{myitemize}{itemize}{1}
\setlist[myitemize]{
    label=\textbullet,
    align=left,
    leftmargin=*,
    nosep,
}

\newlist{myenumerate}{enumerate}{1}
\setlist[myenumerate]{
  label=\arabic*.,
  align=left,
  leftmargin=*,
  nosep,
}

\usepackage{graphicx}
\usepackage{subcaption}
\usepackage{url} 

\usepackage{balance}
\balance

\newcommand{\ignore}[1]{}

\newcommand{\diff}[1]{\textcolor{altgreen}{#1}}

\newmdtheoremenv[
  linewidth=2pt,
  linecolor=black,
  topline=false,
  bottomline=false,
  rightline=false,
  leftline=false, 
  leftmargin=0pt,
  innertopmargin=2pt,
  innerbottommargin=10pt,
  innerrightmargin=10pt,
  innerleftmargin=pt,
  backgroundcolor=gray!10,
  skipabove=\topsep,
  skipbelow=\topsep,
]{definition}{Definition}

\newcommand{\acro}{PEARTS\xspace}
\newcommand{\longacro}{\textit{Provable Execution Architecture for Real-Time Systems}\xspace}

\newcommand{\concept}{RT-PoX\xspace}
\newcommand{\fullconcept}{\underline{R}eal-\underline{T}ime \underline{P}roof \underline{o}f E\underline{x}ecution\xspace\xspace}

\newcommand{\esr}{{\ensuremath{\mathsf{E}\mathsf{S}\mathsf{R}}}\xspace}

\newcommand{\app}{{\ensuremath{\sf{\f}}}\xspace}
\newcommand{\apptask}{{\ensuremath{\sf{S_{App}}}}\xspace}

\newcommand{\napp}{{\ensuremath{\sf{\f_{Adv}}}}\xspace}
\newcommand{\appAll}{{\ensuremath{\sf{S_{All}}}}\xspace}
\newcommand{\rot}{{\ensuremath{\sf{\f_{RoT}}}}\xspace}

\newcommand{\report}{{\ensuremath{\sf{ R}}}\xspace}

\newcommand{\atttask}{{\ensuremath{\mathbb{W}}}\xspace}

\newmdtheoremenv{theo}{Definition}
\newfloat{defbox}{htbp}{lop}[section]
\floatname{defbox}{Definition}

\newcommand{\prv}{{\ensuremath{\sf{Prv}}}\xspace}
\newcommand{\vrf}{{\ensuremath{\sf{Vrf}}}\xspace}
\newcommand{\RA}{{\ensuremath{\sf{ RA}}}\xspace}
\newcommand{\PoX}{{\ensuremath{\sf{ PoX}}}\xspace}
\newcommand{\pox}{{\ensuremath{\sf{ PoX}}}\xspace}

\newcommand{\CFA}{{\ensuremath{\sf{ CFA}}}\xspace}

\newcommand{\CFI}{{\ensuremath{\sf{ CFI}}}\xspace}

\newcommand{\adv}{{\ensuremath{\sf{ Adv}}}\xspace}
\newcommand{\chal}{{\ensuremath{\sf{ Chal}}}\xspace}

\newcommand{\f}{{\ensuremath{\sf{\mathcal F}}}\xspace}

\newcommand{\poxoutput}{{\ensuremath{\Theta}}\xspace}

\definecolor{codegreen}{rgb}{0,0.6,0}
\definecolor{codegray}{rgb}{0.5,0.5,0.5}
\definecolor{codepurple}{rgb}{0.58,0,0.82}
\definecolor{backcolour}{rgb}{0.95,0.95,0.92}

\lstdefinestyle{mystyle}{
    commentstyle=\color{codegreen},
    keywordstyle=\color{blue},
    numberstyle=\tiny\color{codegray},
    stringstyle=\color{codepurple},
    basicstyle=\fontsize{8}{8}\ttfamily,
    frame=single,
    breakatwhitespace=false,         
    breaklines=true,                 
    captionpos=b,                    
    keepspaces=true,                 
    numbers=none,                    
    numbersep=4pt,                  
    showspaces=false,                
    showstringspaces=false,
    showtabs=false,                  
    tabsize=1,
    xleftmargin=0.03\columnwidth,
    xrightmargin=0.03\columnwidth,
}

\newenvironment{propty}[1]{%
  \begin{tcolorbox}[colback=white,
                    colframe=black,
                    title=#1,
                    boxsep=1mm, 
                    left=.5mm,   
                    right=.5mm,  
                    top=.5mm,    
                    bottom=.5mm] 
}{%
  \end{tcolorbox}
}

%% file: pages/0_Abstract.tex
Embedded devices are increasingly ubiquitous and vital, often supporting safety-critical functions. However, due to strict cost and energy constraints, they are typically implemented with  Micro-Controller Units (MCUs) that lack advanced architectural security features. Within this space, recent efforts have created low-cost architectures capable of generating Proofs of Execution (\pox) of software on potentially compromised MCUs. This capability can ensure the integrity of sensor data from the outset, by binding sensed results to an unforgeable cryptographic proof of execution on edge sensor MCUs.
However, the security of existing \pox requires the proven execution to occur atomically (i.e., uninterrupted).  This requirement precludes the application of \pox to (1) time-shared systems, and (2) applications with real-time constraints, creating a direct conflict between execution integrity and the real-time availability needs of several embedded system uses.

In this paper, we formulate a new security goal called \fullconcept (\concept) that retains the integrity guarantees of classic \pox while enabling its application to existing real-time systems. This is achieved by relaxing the atomicity requirement of \pox while dispatching interference attempts from other potentially malicious tasks (or compromised operating systems) executing on the same device. To realize the \concept goal, we develop \longacro (\acro). To the best of our knowledge, \acro is the first \pox system that can be directly deployed alongside a commodity embedded real-time operating system (FreeRTOS). This enables both real-time scheduling and execution integrity guarantees on commodity MCUs. To showcase this capability, we develop a \acro open-source prototype atop FreeRTOS on a single-core ARM Cortex-M33 processor. Based on this prototype, we evaluate and report on \acro security and (modest) overheads.

%% file: pages/999_Official_Main.tex
\section[short]{Introduction}
\label{sec:intro}

The integration of embedded devices across various sectors, including home automation, agricultural technology, wearable gadgets, and smart appliances, has ushered in a multitude of security considerations. While playing a crucial role in enhancing the convenience and connectivity of modern living, embedded devices are characterized by their inherent limitations, which can complicate security efforts. This stems from rudimentary or entirely lacking architectural security features, making them vulnerable to a diverse range of potential threats and vulnerabilities. The importance of addressing security concerns in embedded devices cannot be overstated, especially when deployed in safety-critical environments. In such scenarios, even minor security oversights can yield severe consequences. For instance, the manipulation of sensor data or the disregard of control commands can lead to catastrophic outcomes, ranging from damage to infrastructure to human safety.

To address some of these security concerns, the concept of embedded Remote Attestation (\RA) \cite{smart, vrased, tytan, trustlite, simple, hydra, rata, DAA, Sancus17, scraps, pistis, delegated, reserve, sacha} was proposed to enable remote verification of the integrity and authenticity of embedded device's software. \RA is a (typically hardware-assisted) challenge-response protocol wherein a Prover (\prv) generates cryptographic proof of its internal state to a remote Verifier (\vrf) for examination. Based on the proof, \vrf can decide on the trustworthiness of \prv. Proof of Execution (\PoX)~\cite{mccune2008flicker,apex,asap,rares} extends \RA to bind the data produced by \prv to the correct execution of an intended software routine (e.g., a sensing function), thereby assuring to \vrf integrity of received data from the point when it is first digitized, i.e., the moment physical measurements are first converted into digital data by edge Micro-Controller Units (MCUs). We revisit details of \RA and \pox in Section~\ref{sec:bg}

\subsection{Motivation: On the Conflict Between PoX Integrity \& Real-Time Availability}


Despite its potential to ascertain integrity ``from birth'' of sensor data (even if MCU software is compromised or illegally modified), current \pox methods require the \pox task (here denoted \f) to execute atomically. This ensures that the context (data memory in use by \f), resources (e.g., peripherals in use by \f and their configurations), and timing requirements of \f cannot be tampered with by concurrent applications on \prv. However, atomic execution hinders the application of \PoX to real-time systems, which often rely on Real-Time Operating Systems (RTOS) to ensure strict time-based availability guarantees for various tasks executing on the same MCU. This incompatibility follows naturally from the fact that \PoX must run continuously irrespective of deadlines of other time-critical tasks on the same \prv.

We argue that the reliability of real-time systems simultaneously depends on strong runtime integrity and real-time availability guarantees. 
While \pox and RTOSs may achieve each goal separately, their integration is non-trivial. First, the RTOS should be able to maintain its task scheduling role even when a \PoX instance is active. On the other hand, the authenticity and integrity of the execution being proven must be preserved, despite attempts from other applications or a compromised RTOS to tamper with the \pox state. To make matters more challenging, MCU-based RTOSs (such as FreeRTOS~\cite{freertos}) often operate on single-privilege mode, where all running processes have the same level of access to resources as the RTOS itself.

\subsection{Intended Contributions}
We define \fullconcept (\concept), an advancement over classic \pox to enable preemptive real-time multitasking to securely coexist with \pox. To realize this concept in practice we design \longacro (\acro) by leveraging commodity MCU hardware support from the TrustZone-M~\cite{Armv8_M_TZ_spec} security extension. To the best of our knowledge, this is the first architecture for remotely verifiable execution integrity able to coexist with an unmodified RTOS while retaining both real-time and execution integrity guarantees. 
In sum, our contributions are:

\begin{myitemize}
    \item We define the notion of \concept as well as the necessary requirements to achieve \concept in real-time embedded systems. \concept trades \PoX original requirement of atomic execution for the assurance that interrupting tasks can not tamper with the timing, state, or resources belonging to the interrupted task without \vrf detection.
    \item We design \acro to realize \concept goals. \acro leverages TrustZone-M on single-core ARM MCUs and does not rely on trust assumptions about applications or underlying RTOSs (which typically lack isolation and privilege level separation). \acro respects real-time requirements, as scheduled by the RTOS, even during a PoX. Meanwhile, it also monitors for any actions that could violate the integrity of the PoX task across interrupt cycles. Importantly, \acro is designed not to require changes to the underlying RTOS.
    \item To demonstrate \acro practicality, we develop and evaluate a fully functional open-source prototype~\cite{repo} that runs alongside a well-known and unmodified commodity RTOS (FreeRTOS) deployed on a STM32L552ZE MCU~\cite{stm32l552} featuring a single-core ARM Cortex-M33 processor. Our results show that \concept is obtainable at a relatively small overhead.
\end{myitemize}


As discussed in the rest of this paper, achieving these goals requires addressing several conceptual and technical challenges. First, the single-privilege RTOS model implies a strong adversary (\adv) who may alter both \app and RTOS behaviors.
At the same time, we strive for an approach that does not require modifying the underlying RTOS implementation,
enabling seamless integration with existing settings and use cases. 
To that end, \acro leverages ARM TrustZone-M to create an Elastic Secure Region (\esr) used to monitor \app's context, control flow, and resources and detect interference attempts during a \pox instance (see Section~\ref{sec:details}).
However, as the RTOS resides in the Non-Secure World (see Section~\ref{sec:bg} for TrustZone background), it cannot directly interoperate (e.g., act upon scheduling decisions) with a TrustZone-protected task. Compatibility thus requires the creation of a mechanism (that we name ``shadow'' task) to emulate the (unmodified) RTOS' view of \app as a normal task while dispatching RTOS decisions related to \app into actions to be performed by TrustZone's Secure World.

\section[short]{Background}\label{sec:bg}

\subsection{\RA and \pox}

\begin{figure}
  \centering
    \includegraphics[width=.7\columnwidth]{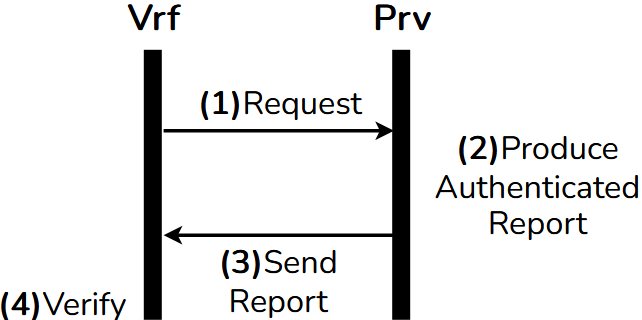}
  \caption{Transcript of \RA/\pox interactions}
  \label{fig:RA}
\end{figure}

\RA is a challenge-response protocol in which \vrf aims to check the software image currently installed on \prv, i.e., the content of \prv's program memory (PMEM). As shown in Figure~\ref{fig:RA}, a typical \RA protocol is performed as follows:

\begin{enumerate}
  \item \vrf requests \RA from \prv by sending a unique cryptographic challenge (\chal).
  \item 
  An \RA root of trust (RoT) within \prv computes an authenticated integrity-ensuring function over PMEM and \chal to produce an authenticated report ($H$).
  \item \prv sends $H$ to \vrf.
  \item \vrf compares $H$ to its expected value.
\end{enumerate}

Step (2) can be implemented using a Message Authentication Code (MAC) or digital signature.
The secret key used in this operation must be securely stored and used by the RoT on \prv to ensure it is inaccessible to any untrusted software. As a result, \RA protocols usually rely on hardware support to implement this RoT on \prv, e.g., from Trusted Execution Environments (TEEs), as in the case of this work.


\PoX augments \RA with evidence that a particular software function within PMEM has been correctly executed and that any claimed outputs were direct outcomes of this timely and authentic execution. This is achieved by modifying step (2) in the \RA interaction with the capability to compute a different report $H_{exec}$ if and only if a function \f, chosen by \vrf and specified along with the request (step 1), is executed in between \prv receiving \chal and computing/sending the authenticated report to \vrf (step 3).

To ensure that the execution of \f is not tampered with by untrusted software on \prv other than \f, the RoT suspends all other applications before calling \f's execution. As \f executes, it produces output \poxoutput. The \pox report is then computed by signing (or MAC-ing) \f's binary, \poxoutput, and \chal. The RoT aborts the process (thus failing to produce $H_{exec}$) if any attempt to interrupt or tamper with \f's execution is made, as interrupting software could tamper with \f ongoing state.
As noted in Section~\ref{sec:intro}, this atomic execution requirement limits \pox practicality due to its direct conflict with real-time requirements of several embedded applications.




\subsection{TrustZone for ARM Cortex-M}\label{subsec:trustzone}

TrustZone-M is a security extension for ARMv8 Cortex-M MCUs, establishing a TEE by partitioning CPU resources into two domains known as the ``Secure" and ``Non-Secure" Worlds. Each of these domains receives dedicated and physically isolated resources, including memory and peripherals.

\textbf{Memory Protection.} To enforce memory isolation, TrustZone-M implements two hardware controllers: the Secure Attribution Unit (SAU) and the Implementation-Defined Attribution Unit (IDAU). SAU grants the system the ability to define security designations (Secure or Non-Secure) to memory segments and allows for dynamic configuration of these designations at runtime. Conversely, the IDAU allows the establishment of security levels for predetermined memory blocks that are defined by the device manufacturer and unmodifiable thereafter. 
%

As Cortex-M devices do not feature virtual memory/Memory Management Units (MMUs), a Memory Protection Unit (MPU) is used to implement access control policies directly -- in physical memory. When TrustZone is present, the MPU is divided into two segments: the Secure MPU (S-MPU) for the Secure World and the Non-Secure MPU (NS-MPU) for the Non-Secure World. Access to the S-MPU configuration registers is only allowed to the Secure World, while the NS-MPU registers are accessible to both worlds.


\textbf{Peripherals Isolation.} Peripherals in Trustzone-M can be categorized into three distinct types based on their security capabilities: non-securable, securable, and TrustZone-aware peripherals. Non-securable peripherals are those that do not support any security features; they are always accessible from both Secure and Non-Secure worlds and cannot be protected by TrustZone mechanisms. Securable peripherals, on the other hand, support basic security features that allow them to be designated as either Secure or Non-Secure. TrustZone-aware peripherals are equipped with intrinsic mechanisms that not only allow them to be designated as Secure or Non-Secure but also enable them to react dynamically to security states or conditions. 

 


\textbf{Interrupt Isolation.} To handle interrupts, the Cortex-M architecture includes a Nested Vectored Interrupt Controller (NVIC). When TrustZone-M is enabled, it supports two Interrupt Vector Tables (IVTs): one for the Secure and one for
 the Non-Secure World.
Thus, each interrupt is individually configured as Secure or Non-Secure. The NVIC includes a register called Interrupt Target Non-Secure (NVIC\_ITNS) used to designate interrupts as belonging to the Secure and Non-Secure Worlds. This register can only be configured by the Secure World. 

 Secure interrupts can be configured to have higher priority than Non-Secure interrupts. When a high-priority Secure interrupt occurs, the CPU pauses Non-Secure operations and redirects execution to the respective Interrupt Service Routine (ISR), implemented in the Secure World. During this process, hardware stores the register context of the Non-Secure World automatically in the Non-Secure stack and changes the processor state to Secure. After handling the interrupt, the ISR updates the program counter (PC) to trigger the {\it Exception Return} routine, restoring the previous state and resuming Non-Secure task execution.


\subsection{Embedded Real-Time Operating Systems and ARM Cortex-M}
\label{sec:background arm-m}



RTOSs are widely used specialized operating systems designed for resource-constrained embedded systems with real-time response requirements~\cite{rtos}. Their core attribute is deterministic behavior, achieved through task prioritization based on predefined deadlines, ensuring the timely and predictable execution of critical tasks.
However, their focus on real-time constraints and simplicity often leaves security practices (e.g., isolation) as a secondary priority.
For instance, widely used RTOSs such as FreeRTOS, are typically used in a single-privileged setup, having user-level and kernel-level code to coexist at the same privilege level without isolation~\cite{khan2023low}.


\textbf{Systick and PendSV.} Typically, embedded RTOSs use a timer to switch active tasks. The timer triggers periodic interrupts, known as ``systicks''. When these interrupts occur, the RTOS scheduler checks the priorities assigned to current tasks. If needed, it switches to execute a different task. In ARM Cortex-M systems, instead of immediately switching tasks during the systick, it is common for the systick ISR to set up another interrupt called PendSV to happen afterward. When no higher priority interrupts are pending, PendSV is activated to perform the task context switch. This ensures precedence to critical interrupt handling over task switching.

\textbf{Stack Pointers.} In bare-metal RTOSs, two distinct stack pointer registers are available: the Process Stack Pointer (PSP) and the Main Stack Pointer (MSP). PSP is primarily used by application threads (tasks) for regular stack operations. MSP is used by the RTOS during boot, initialization, and by ISRs. When TrustZone is present, PSP and MSP are banked across the Secure and Non-Secure worlds. This means that each world maintains its own PSP and MSP registers, allowing for secure and non-secure task stacks to be managed independently 

\textbf{OS Services.} RTOSs typically offer a variety of services for tasks. In single-privilege environments, where conventional system calls for privilege elevation are inexistent, these services are often accessed through direct method/function calls, instead of system call exceptions. Typical RTOS services include dynamic task management for creating and adjusting tasks; scheduling algorithms for task prioritization and timing; synchronization mechanisms such as semaphores and mutexes for coordination; precise timing services using timers; interrupt handling for responding to external stimuli; memory management for optimal resource allocation; inter-task communication for data transfer; and error handling mechanisms to ensure system stability and reliability. 

\section{\concept Definition}

\begin{figure}[t]
    \centering
    \includegraphics[width=\columnwidth]{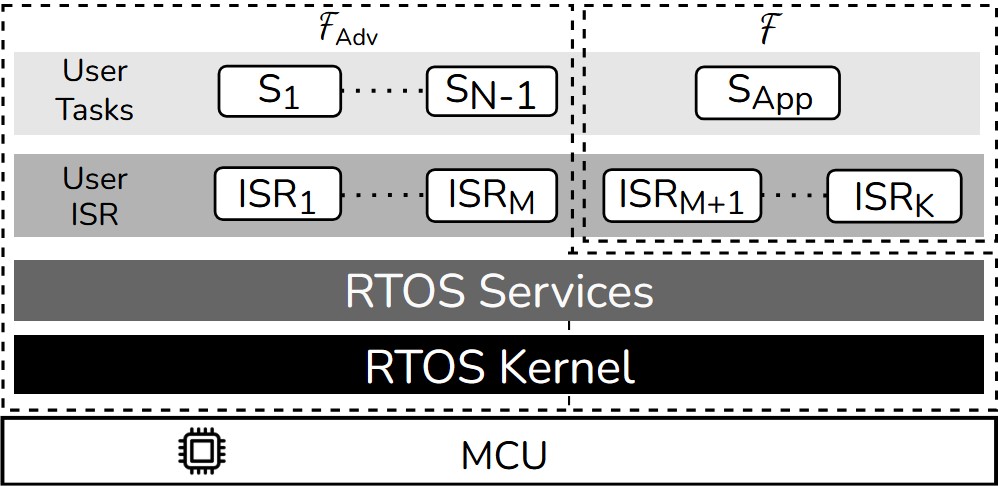}
    \\ \scriptsize ISR = Interrupt Service Routine \normalsize
    \caption{RTOS and software modules on \prv}
    \label{fig:system}
\end{figure}

\subsection{Scope \& Notation}\label{sec:scope}


\concept targets settings that (contrary to classic \pox) depend on an RTOS to schedule multiple tasks on \prv. \vrf wants to assess if an output \poxoutput was produced by the timely execution of a \vrf-defined function \app on a specific \prv, checking that \poxoutput was not corrupted by any other (potentially malicious) software simultaneously installed and running on the same \prv. 



As illustrated in Figure \ref{fig:system}, the RTOS manages and preemptively schedules $N$ software tasks, denoted as a set $\sf{S=\{S_{1},...,S_{N-1}, \apptask\}}$. Along with tasks, \prv implements $K$ \underline{I}nterrupt \underline{S}ervice \underline{R}outines ($ISR$s), denoted $\sf{I=\{ISR_1,..., ISR_{M+1}, ..., ISR_K\}}$. Among tasks running on \prv, \vrf wants to receive a \pox of \apptask. $\apptask$ behavior may depend on inputs obtained through the asynchronous (i.e., upon interrupt triggers) execution of one or more $ISR$s 
$\{\sf{ISR_{M+1}}, ... ,\sf{ISR_K}\} \subset I$. 
A valid execution of \apptask must start from entry point instruction ($s_{entry}$) and conclude at an exit point ($s_{end}$). We refer to \apptask and the code of all $ISRs$ it depends upon as a self-contained functionality \app. We define untrusted code (\napp) to encompass the RTOS and other tasks and ISRs that are not in \app. This separation is also depicted in Figure~\ref{fig:system}.

According to the constraints discussed in Section~\ref{sec:intro}, we consider \prv as low-power single-core MCU with limited program (PMEM) and data (DMEM) memory in the order of dozens of kilobytes. \prv runs software at ``bare-metal'', executing instructions in place (physically from PMEM), and with no 
MMU
to support virtual memory or inter-process isolation.

Furthermore, we define 
\app's execution context $Ctx(\app)$
as all physical memory addresses written and read by \app during its execution, including input arguments given to \app, any global variables used by \app, and any physical peripheral addresses/peripheral configuration addresses \app may use/depend upon (e.g., General Purpose Input/Output (GPIO) and associated configuration registers).


\subsection{\concept Requirements}
\label{subsec:requirements}
Removal of atomic execution introduces significant security challenges as it opens the door to illegal modifications to $Ctx(\app)$ during \app execution. Such incidents can stem from interrupts that result in \napp controlling \prv state while \app execution is paused (but not completed). The same applies when \app demands RTOS services and issues system calls.

It follows that \app execution integrity depends on \napp non-interference with $Ctx(\app)$ while \app executes, i.e:
\begin{equation}\label{eq:non-int}
    Ctx(\app) \cap Ctx(\napp) = \emptyset,  
\end{equation}
where $\cap$ denotes interference, i.e., Equation~\ref{eq:non-int} asserts that no physical memory address accessed by \app during its execution is modified by \napp.

Despite non-interference stipulated in Equation~\ref{eq:non-int}, timing attacks \cite{bechtel2019denial,mahfouzi2019butterfly,wittenmark1995timing} may still affect critically timed behavior in \app. 
Therefore, \vrf must be able to assess any timing requirements that pertain to the correctness of \app. The root of the timing issue lies in the ability of \napp to artificially delay the resumption of \app. Therefore, \app interrupt locations (i.e., at which instruction \app is paused) and periods (for how many clock cycles) must be observable by \vrf in the \pox result.

Aside from timing, \vrf must be informed of the nature of transitions between the \app and \napp to define what constitutes a valid interrupt delay. This is because different interrupts/system calls or the same system calls with different arguments have varied timing constraints, e.g., a task ``delay'' system call (namely vTaskDelay in FreeRTOS) with argument ``10 seconds'' should take 10 seconds before resuming the caller task. Earlier resumption may cause the caller to access a resource that is not yet ready. Late resumption may result in missed sensing/actuation deadlines.



Finally, \textbf{execution control flow path attacks} can be leveraged by \napp to divert the correct order in which instructions of \app should execute. This is possible without changing memory in use by \app or its timing, by simply jumping back to the incorrect instruction address of \app when resuming from an interrupt or a system call.


Given the aforementioned challenges, we stipulate that a secure \concept must guarantee the following Properties~1-5 to support \app integrity irrespective of \napp's control of \prv state through interrupt periods within \app execution.

\begin{propty}{\textit{\textbf{Property 1 -}}\label{prop:1}
\textbf{Ephemeral Immutability.}}
The binary of \app (in \prv's PMEM) must be immutable while \app executes and in between its execution and measurement (i.e, the latter part of the \pox procedure -- recall Section~\ref{sec:bg}).
\end{propty}


Without Property 1, \napp could change the binary of \app (which by definition also includes the code implementing ISRs \app depends upon). Clearly, this would modify \app's behavior. Note that simply disabling all runtime modifications to PMEM in a single privilege system impedes remote software updates~\cite{rata}, which is impractical in many settings. Instead, \concept must ensure that \app binary remains consistent (with the hash value reported to \vrf) during its execution, allowing PMEM modifications otherwise.

\begin{propty}{\textit{\textbf{Property 2 -}}\label{prop:2}
\textbf{Execution Flow Integrity.}}
Interrupts, system calls, or execution of \napp must not affect the integrity of \app's control flow. Additionally, \app must start from its valid entry point ($s_{entry}$) and terminate at its valid exit ($s_{exit}$).
\end{propty}



Per Property 2, \concept requires that untrusted code execution due to transfers from \app to \napp — such as interrupts, system calls, or task switches — must not affect the integrity of \app's execution flow when \app execution resumes.



\begin{propty}{\textit{\textbf{Property 3 -}}\label{prop:3}
    \textbf{Context Integrity.}}
    
    Data in use by \app must not be subject to influence from \napp. This includes physical memory used by \app in DMEM, register contents, and data inputs obtained from I/O peripherals.
\end{propty}
%


Property 3 ensures that \app does not rely on data provided or modified by \napp, including any changes to peripheral configurations (e.g., timer or GPIO resolution) that could indirectly affect \app's I/O values obtained by \app. 
We note, however, that data inputs used by \app can be influenced by external temporal/physical factors (which are independent of \prv's internal software state). For example, a physical temperature change could (and should) affect the data inputs of \app if \app is for instance a temperature-based application performing a temperature read.


\begin{propty}{\textbf{\textit{Property 4 -}}\label{prop:4}
    \textbf{Timing Integrity.}}
    \concept should provide \vrf with information about any delays to \app caused by \napp execution periods (due to interrupts and system calls).
\end{propty}


Property 4 allows \vrf to decide if any delays caused are still acceptable for the functionality implemented by \app.

\begin{propty}{\textbf{\textit{Property 5 -}}\label{prop:5}
\textbf{RTOS Availability.}}
\concept must co-exist and not interfere with RTOS duties, e.g., task scheduling, timing, and resource allocation.
\end{propty}
\vspace{0.5em}

Property 5 is the {\it raison d'$\hat{e}$tre} of real-time systems, where the predictability and reliability of task execution time are vital. It is also the main motivation of \concept over classic \pox. \concept enables the co-existence of PoX and a Non-Secure World RTOS responsible for scheduling. This maintains the RTOS outside the PoX trusted computing base (TCB).
We highlight that \concept focus is not to guarantee proper scheduling/availability despite a compromised RTOS. The latter is the focus of mechanisms supporting trusted scheduling (e.g.,~\cite{proactive3,aion,garota,wang2022rt}). Appendix C further elaborates on this difference.




\begin{figure*}[t]
    \centering
    \includegraphics[width=.95\linewidth]{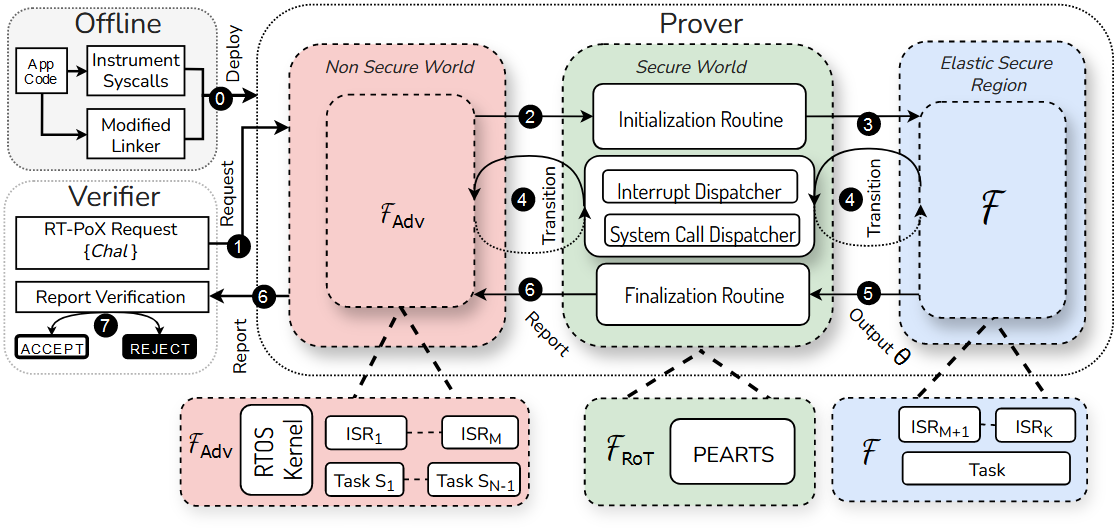}
    \caption{\concept steps with \acro.}
    \label{fig:system2}
\end{figure*}
    \section{\acro Overview}

\acro realizes \concept on an off-the-shelf embedded RTOS (FreeRTOS) running on commodity MCUs (ARM Cortex-M) and does not require RTOS modifications.
Toward this goal, it leverages ARM TrustZone-M to establish a secure environment that monitors the execution of \app during a \pox, enforcing \concept properties.

\subsection[short]{System Assumptions}

\noindent\textbf{Device.} \prv consists of a resource-constrained single-core, bare-metal MCU (as described in Section~\ref{sec:scope}) equipped with TrustZone-M (recall Section~\ref{sec:bg} for TrustZone-M details).

\vspace{0.5em}\noindent\textbf{RTOS.} We model the RTOS after the default structure of FreeRTOS, given its widespread adoption in practical settings \cite{marketStudyEmbedded}. It employs systick-based interrupts for time tracking and to decide when a context switch is needed. 
The PendSV ISR implements context switches. 
Tasks and the RTOS run within the same privilege level (no privilege separation/isolation)~\cite{khan2023low}.

\vspace{0.5em}\noindent \textbf{Software Architecture}. The RTOS and tasks it oversees (including \app) are housed in \prv's Non-Secure World. 
\acro RoT implementation, denoted \rot, resides in the Secure World and has exclusive access to a secret key ($sk$). The corresponding public key ($pk$) is securely provisioned to \vrf before \prv's deployment.
%

\textit{\textbf{Remark.} We note that the trivial approach of implementing \app within the Secure World (thus isolated from the Non-Secure World) does not achieve \concept. Such an approach does not allow \concept of multiple, arbitrary (\vrf-configurable) applications, as it would fix \app to code shipped as part of the Secure World, at device deployment time. Furthermore, it would mix the TCB of \acro's  RoT with the untrusted code of multiple applications that \vrf may want to execute provably (within the Secure World, in this hypothetical case).}


\subsection{Adversary Model}

The Adversary (\adv) is assumed to have full control over \prv's Non-Secure World and can alter its code (e.g., via code injection attacks) and exploit vulnerabilities to cause control flow hijacks and code reuse attacks. The RTOS in the Non-Secure World is not considered intently malicious, but rather untrustworthy (i.e., could contain exploitable bugs).
\adv can also attempt to manipulate \app (such an attempt should be detected by a secure \concept -- recall Property 1). \adv can change Non-Secure World interrupt configurations to trigger ISRs at will and can alter or corrupt ISR implementations.
However, given TrustZone-M protections, \adv cannot access or tamper with code or data in the Secure World. Similarly, it cannot deactivate or circumvent the TrustZone-M hardware-enforced access control rules and assurances.
Physical and hardware-modifying attacks are out-of-scope, as they require orthogonal tamper resistance measures~\cite{obermaier2018past}.



\subsection{Workflow}\label{sec:workflow}

At \app compilation time, \app binary is generated with a custom linker to make its code self-contained/independent from \napp. 
\acro supports but does not mandate the assignment of ISRs as a part of \app. If \app functionality relies on ISRs, relevant ISRs (and their dependencies) must be linked within \app's binary as a part of \app implementation itself; otherwise, they can be excluded from \app.
\app system calls are also instrumented, as depicted in \circ{0} from Figure~\ref{fig:system2}.
The instrumentation of \app system calls is used to ensure that they are intercepted and adequately dispatched by \acro's RoT. \acro system call dispatching ensures that system call timing requirements can be verified irrespective of \napp behavior.

A \acro instance begins when \vrf sends an \concept request (containing challenge \chal)  to \prv, requesting a timely execution of \app, as shown in \circ{1}.
Upon receiving the request, \prv must call \rot to start the \acro Initialization Routine \circ{2}. This routine sets up the initial transition from \napp to \app. This includes configuring a so-called Elastic Secure Region (\esr). \esr is a memory region in the Non-Secure world used to isolate \app memory from \napp. Additionally, the initialization routine sets up all interrupts that do not belong to \app, to be trapped in the Secure World. This ensures that any eventual illegal memory accesses and control flow transitions between \app and \napp are observable by \rot. 
Furthermore, all peripherals are initially configured to be inaccessible to the Non-Secure World (including both \napp and \app). This traps \app attempts to access peripherals to \rot, allowing \rot to mark them as ``in use'' by \app and detect subsequent interfering accesses by \napp.


Once the setup is complete, \rot launches \app execution in \esr by switching the CPU mode to Non-Secure and jumping to $s_{entry}$ (step \circ{3}). To support other real-time operations on \prv, \app execution may be sporadically paused to execute \napp (e.g., due to \app-issued system calls or \napp interrupts/context switches).
Nonetheless, due to the traps created in the Initialization Routine, \rot can monitor these transitions (step \circ{4}) to detect control/data-flow violations as well as to record the duration of each transition. Once \app concludes at $s_{end}$, producing the output \poxoutput, it returns to \rot, as shown in \circ{5}. 

During \napp execution, any resources not exclusively used by \app are still freely accessible, allowing \napp to operate normally without being blocked by \acro.
The access to resources in use by \app is monitored by \rot to ensure non-interference by \napp.
A so-called {\bf interference exception} is triggered if \napp attempts to: (i) access one of \app's exclusive resources (\app code/data or peripherals) or (ii) execute code within \esr (e.g., directly jump to \esr without going through \rot).
To preserve the RTOS availability, \rot records the exception and releases the resource to \napp. 
The recorded exception is later included in the \concept report (\report) along with \poxoutput and \app measurement, which are all signed by \rot and sent to \vrf (in \circ{6}).
Upon receiving \report, in \circ{7}, \vrf considers \poxoutput trustworthy only if: (i) the signature is valid; (ii) no interference exceptions are reported\footnote{Optionally, \vrf may choose to analyze the context of specific interferences to judge if they could have affected \app outcome.}; (iii) all reported transitions between \app $\leftrightarrow$ \napp occur within acceptable duration (according to \app requirements).

\subsection{\acro Guarantees at a High Level}


Before delving into \acro details in Section~\ref{sec:details}, we here discuss the high-level rationale behind the workflow discussed in Section~\ref{sec:workflow} in guaranteeing the properties introduced in Section~\ref{subsec:requirements}.

\textbf{Property 1} (\app immutability) is attained by having \app code in \esr. 
Whenever a transition to \napp occurs during \app execution, \rot takes over the control to configure \esr (via the SAU TrustZone controller, recall Section~\ref{sec:bg}) as a Secure Region before switching to \napp. With this configuration, any \napp attempts to access \esr will cause a fault; this fault redirects the control to \rot, which saves such attempts to include them in \report that will later be sent to \vrf. \rot reverts \esr back to a Non-Secure Region when transitioning back to \app execution. This approach also ensures that \app TCB and \rot TCB are isolated from each other (as \app does not execute in the Secure World). Since \napp cannot modify \app code (located inside \esr) without \rot detection, \textbf{Property 1} is fulfilled.

For \textbf{Property 2} (execution flow integrity), \rot stores the proper return address before dispatching interrupts/system calls to \napp and resumes \app at this address after \napp execution concludes. Since the return address is stored in the Secure World, it cannot be manipulated by \napp. \napp may also attempt to cause execution flow attacks by directly jumping to the middle of \app. However, as \app code resides in \esr (inaccessible to \napp) such attempts lead to an interference exception which is therefore detected and included on the report to \vrf by \rot.

Recall that 
$Ctx(\app)$ 
consists of all memory regions allocated or used during \app execution, including its DMEM, peripheral addresses, and peripheral configuration addresses.
\acro achieves \textbf{Property 3} (Data Integrity) using three methods. 
First, in addition to \app's code, \rot makes \app's DMEM part of \esr.
Doing so prevents illegal access to \app's runtime data during \napp execution.
Second, \rot configures the peripherals used by \app to be inaccessible by the Non-secure World while \napp executes.
This prevents \napp code from altering peripheral configurations or stealing peripheral data meant for \app without detection and logging by \rot.
Third, \app system calls are instrumented with trampolines to the System Call Dispatcher in \rot. 
This causes \rot to intercept them before redirection to the RTOS, allowing \acro to monitor RTOS's response times and log timing manipulations.

\textbf{Property 4} (Timing Integrity) is satisfied in \acro by \rot's implementation always recording each operation (i.e., system calls, returns, interrupts, or returns from interrupts) that triggers a transition during \app execution.
Also, \rot logs the timestamp of each transition using a dedicated secure clock that is only accessible/configurable to/by \rot within the Secure World.

By design, \app provable execution is interruptible while leveraging aforementioned interference exceptions to detect and log (but not block) any external interference. Therefore, an RTOS can still fulfill its duties according to any underlying real-time requirements, realizing \textbf{Property 5} (RTOS Availability).

\label{sec:control_flow_monitor}



\section{\acro Design in Detail}\label{sec:details}

\subsection{Interference Exceptions}\label{sec:interference}


This section explains the resource isolation scheme to enable exception-based monitoring (Section~\ref{subsec:res-iso}) and details how each type of interference exception is handled and reported in \acro reports (Section~\ref{subsec: monitoring violations}). Interference exceptions are active whenever an \concept instance exists in the system, i.e., in between \acro Initialization (Section~\ref{sec:init}) and Finalization (Section~\ref{sec:report}) routines.

\subsubsection{Resource Isolation Scheme}\label{subsec:res-iso}~\\


\noindent\textbf{Code \& Data Isolation.}
The Non-Secure memory (including PMEM and DMEM) is partitioned into two regions, \esr and the rest of the system memory, each containing its own code and data sections.
\esr is used for storing \app and the rest is for \napp. 
During a \acro protocol instance, whenever a transition happens from \app to \napp, \rot marks the \esr as a Secure Region by modifying the SAU controller configuration.
As a result, any subsequent attempts by \napp to access (read, write, or execute) \esr will trigger an interference exception, notifying \rot about these attempts. Conversely, when \napp returns the control to \app, \rot sets \esr as a Non-Secure Region, re-enabling execution and access to \app in the Normal World.


\vspace{0.5em}\noindent\textbf{Peripheral Isolation.} 
At the beginning of \acro protocol, \rot configures all peripherals to the Secure state.
This restricts direct access to all peripherals from Non-Secure World (including \app) during \app execution.
When \app needs to access a certain peripheral, it must request access to \rot.
Upon receiving it, \rot switches the peripheral to the Non-Secure state and associates it with \app. 
Whenever a transition $\app \rightarrow \napp$ occurs, \rot configures peripherals in use by \app as belonging to Secure-World and other peripherals as belonging to Normal World.
As a result, \napp attempts to access \app peripherals lead to interference exceptions, thereby notifying \rot of the attempted access.

\textit{\textbf{Remark.} The secure configuration of peripherals does not prevent \napp from accessing these resources. Instead, it is leveraged to trigger \rot to quickly log the peripheral access before allowing it (see details below). This ensures that the \concept report informs \vrf of any potential interference of \napp on peripherals that \app may depend upon during execution.}

%

\subsubsection{Interference Exception Handling}
\label{subsec: monitoring violations}~\\

When an interference exception is triggered, \rot examines the source address that caused the fault to determine its origin. A flag corresponding to the type of fault cause is added to the \concept exception log ($\mathbb{E}log$). $\mathbb{E}log$ is subsequently included in the \concept report. If the fault was caused by an access to \esr, \acro deactivates the \esr, re-configuring it as a Non-Secure region to allow the access (as this could be a time-critical access to a resource in use by \app).
If the fault resulted from peripheral access, \acro re-configures the peripheral to be part of the Non-Secure world after logging the access. By allowing \napp to continue operation after a fault, \acro avoids disrupting other potentially critical tasks in \prv and allows \app to resume its execution afterward. At the same time, it ensures that any interference is logged in the \concept report that is sent to \vrf after \app execution.

%
To track time delays that \napp interrupts might introduce to \app's execution we define a tuple to be measured for each transition \napp $\leftrightarrow$ \app (in both directions). This tuple captures the execution timestamp $t$ of the context switch and is structured as

\begin{equation}\label{eq:trans}
    e = [Op, Src, Dst, Args, t]
\end{equation}
where:
\begin{myitemize}
\item $Op$ is the operation initiating the transition between \app and \napp (a particular interrupt or system call);
\item $Src$ is the last instruction address executed before the transition;
\item $Dst$ is the jump-to destination address of the transition; and
\item $Args$ contains relevant arguments in the transition (e.g., system calls parameters, if applicable).
\end{myitemize}
The sequence $\mathbb{T}log = [e_0, ..., e_e]$ (where each $e_i$ has the form defined in Equation~\ref{eq:trans}) contains information of all transitions during \app execution and is also included in the \concept report sent to \vrf.
 

\subsection{Initialization Routine}\label{sec:init} 


Upon receiving an \concept request from \vrf, RTOS creates a task, associated with the \concept execution, that triggers the \rot to start \acro Initialization Routine, performing the following:

\begin{myitemize}
    \item \textbf{Memory Allocation}. \rot reserves memory segments within \esr for \app data allocation. This includes moving \app stack pointers (PSP and MSP, recall from \ref{sec:background arm-m}) and heap structures to fall within that range.
    
    \item \textbf{Peripheral Configurations}. All the peripherals are set to be accessible only by the Secure World. This enables the exception-based interference monitoring described in Section \ref{sec:interference}.
    \item \textbf{DMA Configuration.} \rot checks if any DMA channel is configured to write within \esr. If so, a record is added to $\mathbb{E}log$ detailing the address range and peripheral(s) involved in DMA writing. 

    \item \textbf{\app Binary Measurement.} \rot hashes \esr PMEM and the NS-IVT (which contains code pointers to the ISRs called upon each possible interrupt trigger, including the ISRs belonging to \app). This result -- $H(\app)$ -- is later included in the \concept report.
    \item \textbf{Interrupt Dispatchers.} \rot configures all interrupts that are not associated with \app to be trapped by the Secure World. This setup intercepts any transitions between \app and \napp triggered by interrupts to ensure that they are securely dispatched back to their original ISRs in the Non-Secure World (see dispatching details in Section~\ref{sec:dispatch}).
    \item \textbf{Secure Timer.} \rot reserves a hardware timer to timestamp transitions between \app and \napp, monitoring the time taken by \napp to resume \app. 
\end{myitemize}

\subsection{Interrupt Dispatcher} 
\label{sec:dispatch}
Recall that \acro redirect interrupts to the Secure World so that \rot can configure the system to enforce non-interference with \app before allowing Non-Secure ISR handling.
To redirect interrupts, \acro configures them as Secure.
This implies that the S-IVT is used to locate the ISR responsible for handling each trigger (instead of NS-IVT). This is accomplished by setting the corresponding bits in the NVIC\_ITNS registers (recall Section~\ref{subsec:trustzone}) to 0 for all interrupts. \acro maintains two versions of the NVIC\_ITNS: one for when \app is active and another for when \napp is active.
When \app is active, the interrupts not related to \app are marked as Secure in the NVIC\_ITNS while the \app-related ones are marked as Non-Secure. Conversely, when \napp is active, all interrupts from \app are set as Secure and all others return to the original security state. This dynamically adjusts the security settings based on the active application to maintain operational integrity.

Note that the assignment of interrupts to \app should be based on whether \app's functionality depends on the ISR execution. If \app does not rely on any interrupts, all interrupts should be assigned to \napp.

\subsection{RTOS System Calls Within \app}
\label{subsec: rtos service dispatcher}

\app should be self-contained to include all code its functionality depends upon.
However, to maximize \prv utilization by other simultaneous tasks, \app must still use RTOS system calls related to scheduling and time management. For instance, if \app must ``sleep'' for 10 seconds, it should be able to use the RTOS system calls to sleep. In that way, the RTOS would be notified of the 10-second slot made available by \app and potentially allocate it to other tasks, increasing system utilization.
However, reliance on system calls implemented by the potentially compromised RTOS may result in violations to \app's execution integrity, e.g., if instead of waking up \app after the expected 10 seconds, \napp were to hold control for a shorter or longer period.

\acro allows \app to make RTOS system calls for functions that involve time management, scheduling, and synchronization. However, instead of directly handing control to the RTOS, \acro instruments the aforementioned system calls to trigger \rot to start its own measurement of the time taken to resume \app. This is achieved using the \rot-dedicated Secure Timer discussed in Section~\ref{sec:init}. The secure time measurements are used to construct $\mathbb{T}log = [e_0, ..., e_e]$ (as discussed in Section~\ref{sec:interference}). At the same time, this enables the use of the default RTOS system calls for time management and scheduling of \app together with other tasks in the system. Appendix B lists system calls supported in \acro prototype (along with related security considerations).
\subsection{Auxiliary \rot Services}
\label{sec:system_calls}
\textbf{Heap Memory Allocation.}
In real-time embedded systems, dynamic memory allocation is often avoided due to overhead and, in some cases, unpredictability.
In situations where dynamic allocation is required, the memory allocation process is handled by an RTOS service. In our scenario, however, \app cannot yield control over its memory to \napp. Thus, when \app uses dynamic allocation, a separate instance of the memory allocator is implemented within the \esr.
When dynamic memory allocation is required by \app, \rot assigns space in \esr designated as the heap during the Initialization Routine.

\textbf{DMA Management.} \label{subsec:DMA}
CPU-independent writes by DMA might attempt to modify \esr. To mitigate this, \acro revokes DMA permissions from the Non-Secure World and provides an API that allows the Non-Secure World to request DMA operation configuration to \rot. This allows \rot to log DMA attempts that interfere with $Ctx(\app)$ to $\mathbb{E}log$.

\textbf{Self-Timing.} \app can perform secure time measurements during its execution through a request to a time function implemented by \rot. This is implemented using the same reserved secure timer discussed in Section~\ref{sec:init}, also used to time \napp execution periods when \app is interrupted.




\subsection{Finalization Routine and Verification} \label{sec:report}

After \app execution concludes, the Finalization Routine is triggered. This routine restores the system's configuration to its state prior to the \concept process. It frees DMA and peripherals for the Non-Secure World, sets \esr as Non-Secure, clears runtime memory, and restores interrupt configurations. Finally, it generates the attestation report \report.
 
The \concept report $\report = \{ \sigma_{\app},\mathbb{E}log,\mathbb{T}log,\poxoutput\}$ contains information needed by \vrf to assess the integrity of \app's execution on \prv. $\mathbb{E}log$ contains all external (\napp) accesses that interfered with \app's context 
 during its execution. $\mathbb{T}log$ contains the times and context switch information of all interrupts and system calls during \app's execution. \poxoutput is the output of \app execution. $\sigma_{\app}$ is the signature
\begin{equation*}\small
    \sigma_{\app}=Sign_{sk}(H(\app)||\chal||\mathbb{E}log||\mathbb{T}log||\poxoutput)
\end{equation*}
computed using the Secure World's secret key $sk$. Therefore, it authenticates \app's binary, $\mathbb{E}log$, $\mathbb{T}log$ and \poxoutput, and proves the freshness of $\report$ through \chal.
Importantly, $H(\app)$ and $\chal$ are committed to an \concept instance by the Initialization Routine before \app execution. \app (in \esr) is immutable thereafter. On every \concept instance, $\sigma_{\app}$ is computed on the same $H(\app)$ and $\chal$ committed during initialization and only after \app execution reaches its last instruction ($s_{exit}$), indicating \app's execution has completed.




After authenticating $\sigma_{\app}$, checking $H(\app)$, and checking $\chal$, \vrf can examine $\mathbb{T}log$ for adherence to application-specific timing constraints. Similarly, it can inspect $\mathbb{E}log$ to verify that \app context was not tampered with.

\section[short]{Security Analysis}

\adv's goal is to convince \vrf that \app executed correctly on \prv (optionally producing some $\poxoutput$ output) when no such execution occurred, when its context or timing constraints were tampered with, or when $\poxoutput$ is not a direct result of said execution. Towards this goal \adv might attempt to perform:

\begin{myitemize}
\item \textbf{\app binary modifications.} \adv might change \app binary prior to the initiation of \concept process. However, as $H(\app)$ is computed during initiation, any such modification would result in a mismatch with the code hash expected by \vrf. \adv could also attempt to alter the \app binary during the \concept process (after $H(\app)$ computation). However, due to SAU-enforced protection to \esr, this attempt is detected by \rot and thus by \vrf.

\item \textbf{\app execution flow manipulation.} To manipulate the execution flow of \app during an \concept process, 
\adv must return from \napp to an incorrect instruction in \app.
Directly jumping to an arbitrary address within \app is infeasible because during \napp execution \esr is set as a Secure Region by SAU (hence not executable from the Non-Secure World). 
Alternatively, \adv might attempt to change intermediate control flow-related data in \app's stack. 
This also fails because \app's data is also within \esr. 
Finally, \adv could try to modify the interrupt vector tables to change the address executed upon interrupts during \app execution. However, \acro checks the integrity of the IVT addresses related to \app before \app execution and maintains a copy of the IVT in the \esr (which is always checked upon resuming \app execution). 

\item \textbf{\app data corruption.} As previously discussed, \adv is unable to directly modify the data within \app due to SAU memory protections to \esr. Consequently, \adv may attempt alternative strategies to modify $Ctx$(\app). One such strategy involves altering the configurations of peripherals used by \app. This alteration can change the expected behavior of the peripheral and affect the data that \app subsequently reads from it. However, when \napp is active, all peripherals accesses are trapped by the Secure World and logged, preventing \adv from stealthily modifying the peripheral registers/configurations. %
%
Alternatively, \adv can attempt to use DMA to corrupt \app data.
This is not feasible because, during the \concept process, all DMA channels are configured securely by \rot. \acro also checks for any pre-configured DMA channel targeting \esr during the \concept initialization.

\item \textbf{Timing attacks.} \adv may alter response times of time-sensitive operations within \app by interfering with their schedule. \adv could achieve this by either delaying or prematurely advancing the schedule of operations within \app when the schedule is handled by the RTOS. However, \acro independently logs times and associated context of all $\app \leftrightarrow \napp$ transitions using a dedicated timer only accessible in the Secure World. Therefore, timing attacks are detectable by \vrf.

\item \textbf{\concept report forgery.} Given \adv's inability to perform the aforementioned actions, its last resort is to forge the \concept report. This in turn requires forging signature $\sigma_{\app}$ which is computationally infeasible as long as $sk$ is unknown to \adv and the underlying signature primitive is existentially unforgeable. In turn, $sk$ is stored in the Secure World and only accessible to \rot.

\end{myitemize}





\section{Implementation Details}



\subsection{Interfacing the RTOS and \acro}




Housing \app in \esr prevents the RTOS from directly managing it as a normal task in the Non-Secure World. Doing so would raise a security exception (due to TrustZone's world separation) during context switches or system calls.
To allow \acro to run on an unmodified RTOS, a ``shadow task'' is created to mimic \app's presence as a standard RTOS-managed task.
The RTOS creates the Shadow Task in the Non-Secure World normally, using its standard task creation API (e.g., \textit{xTaskCreate} in FreeRTOS).
Shadow tasks have fixed behavior and interface the unmodified RTOS with \rot.
Once the Shadow Task starts executing (as scheduled by the RTOS) it calls \rot to initiate/switch/resume \app.

The Shadow Task also mediates system calls originating from \app. \app system calls are instrumented to yield control to \rot. \rot then changes the context to \napp and directs the execution to the Shadow Task which triggers the respective system call in the RTOS, in the Non-Secure World. 
See Appendix A for implementation details on Shadow Tasks.

\subsection{Context Switches}


\textbf{Switching \app $\rightarrow$ \napp}: 
During this context switch, \esr is configured as a Secure Region, and the peripherals used by \app are set as Secure while all other peripherals are set as Non-Secure. Register context (including MSP and PSP) is saved to the Secure World, and MSP and PSP are replaced with those belonging to \napp to prepare \napp context and stack. The NS-IVT is configured as described in Section~\ref{sec:dispatch}. A new entry is recorded in $\mathbb{T}log$ to log the transition, interrupts are re-enabled, and control is passed to \napp.

\textbf{Switching \napp $\rightarrow$ \app}: Per \acro enforced properties, \rot receives control to initiate this switch. \rot marks \esr as Non-Secure. Next, \app peripherals are reconfigured as Non-Secure, while the remaining peripherals are set to Secure.
The register context of \napp is saved and the \app's register context and the NS-IVT are restored. This transition is logged in $\mathbb{T}log$. Finally, the switch concludes by re-enabling interrupts and resuming \app.



\subsection{Dispatching Banked Interrupts}

When TrustZone is active, certain interrupts (such as systick) are banked between Secure and Non-Secure modes. Thus, they are signaled separately for each world. 
This separation prevents the direct configuration of a banked Non-Secure interrupt to be handled by \acro dispatcher in the Secure World. To address this, \acro keeps a copy of the NS-IVT within \esr, where the addresses of the ISRs for these banked interrupts are replaced with fixed reserved Secure World addresses. When the NS-IVT copy is accessed as part of the Non-Secure World interrupt handling process, any trigger of the banked interrupt will also trigger a Secure World exception. The exception implementation allows the Secure World to verify if the fault originated from a systick exception attempting to access the associated secure address. If so, the fault is redirected to the interrupt dispatcher and then redirected by the dispatcher to the original systick ISR in the Non-Secure World.

\section{Prototype \& Evaluation}


We developed a proof-of-concept prototype of \acro on a NUCLEO-L552ZE-Q development board, which features an STM32L552ZE MCU. This MCU is built on the ARM Cortex-M33 (v8) architecture, operates at 110 MHz, and supports Arm TrustZone-M. \acro functions in the Secure World are implemented in \texttt{C}. SHA224 and HMAC cryptographic operations leverage MCU built-in hardware accelerators and are only accessible by the Secure World. The Non-Secure World runs a standard version of FreeRTOS, with its API interfaces specified in CMSIS-RTOS2 \cite{arm-cmsis}. \vrf is implemented in Python to communicate with \prv, initiate \concept requests, and verify responses. \acro prototype implementation is publicly available in \cite{repo}.
We evaluate \acro prototype regarding its latency, memory, and runtime overheads.
We also analyze \acro's impact on real applications.


\subsection{Runtime Overhead}

\subsubsection{Interrupt Dispatcher Latency}~\\

We start by examining the latency overhead of \acro's Interrupt Dispatcher on \prv. Recall that when \app is active, \rot dispatches \napp interrupts and returns from interrupts, increasing the latency of interrupt handling. This increase in latency can impact the application's time budget. The degree of this impact depends linearly on the interrupt frequency in \napp.

We assess the latency associated with both types of transitions in \acro: $\app \rightarrow \napp$ (interrupt) and $\napp \rightarrow \app$ (return from interrupt, a.k.a. interrupt backtrip). We consider the following metrics:

\begin{myitemize}
    \item \textbf{Interrupt Latency}: The time elapsed from the triggering of an interrupt to the start of its respective ISR execution.
    \item  \textbf{Interrupt Latency Back-trip}: The time from the completion of the ISR execution to the resumption of the interrupted task.
\end{myitemize}


Results are summarized in Table \ref{table:latency}. In all cases, the time unit reported is the number of CPU cycles taken (at 110 MHz, one CPU cycle is $\approx$ 9 nanoseconds). Each measurement was conducted 1000 times, covering all possible execution paths of the interrupt dispatcher, to ensure consistency. During tests, we ensured that no other interrupts occurred between the start and end of the measurements. For comparative analysis, we also measured these components with \acro deactivated, thus establishing a baseline to quantify the overhead introduced by the dispatcher.  

 The analysis of interrupts for \app $\rightarrow$ \napp shows that the  maximum interrupt latency was  102 CPU cycles. Similarly, the  maximum interrupt backtrip latency was 85 CPU cycles. For $\napp \rightarrow \app$, the observed maximum interrupt latency was about 102 CPU cycles, and the maximum interrupt backtrip latency recorded was 84 CPU cycles. While transitions from \app to \napp will always frequently occur due to the scheduler interrupt associated with \napp, transitions from \napp to \app depends more on specific application needs.

\subsubsection{Per-Module Runtime Breakdown}~\\


 \begin{table}[t]
\centering
\caption{Context Switch Latency Introduced by \acro}
\begin{tabular}{l|cc}
\hline
\multicolumn{1}{c|}{}                                 & \multicolumn{2}{c}{\textbf{CPU Cycles}} \\
\multicolumn{1}{c|}{\multirow{-2}{*}{\textbf{Event}}} & \textbf{Max}       & \textbf{Min}       \\ \hline
$\app\rightarrow\napp$ Interrupt                                    & 119                 & 117                 \\
\rowcolor[HTML]{EFEFEF} 
$\app\rightarrow\napp$ Interrupt Backtrip                  & 85             & 85             \\
$\napp\rightarrow\app$ Interrupt                    & 102             & 101             \\
\rowcolor[HTML]{EFEFEF} 
$\napp\rightarrow\app$ Interrupt      Backtrip                           & 84                 & 83                 \\
\hline
\end{tabular}
\label{table:latency}
\end{table}

Table \ref{table:latency extra} shows the runtimes of various operations in \acro that could impact the performance of \prv.

To quantify \acro Initialization Routine's impact, we measured the time taken between RTOS' call to \app's shadow task until \app's \concept execution starting. The maximum initialization latency observed was
around 8.5K CPU cycles ($\approx0.08$ milliseconds). The runtime introduced during the initialization routine is attributed to both the required configuration of hardware controls and the computation of $H(\app)$ (using SHA224). Additionally, we measure the Finalization Routine latency, which involves generating report \texttt{R} (implemented with HMAC) and reverting \prv to its pre-\pox configuration. The maximum finalization routine runtime recorded was 14.3K CPU cycles ($\approx0.14$ milliseconds) when \texttt{R} size was around 2KBytes due to a large number of $Tlog$ entries. As a lower bound, when $Tlog$ is empty, the runtime measured was around 3.2K CPU cycles ($\approx0.03$ milliseconds).
Note that the initialization and finalization routines are executed once per \concept instance, before and after \app execution, respectively. Therefore, they do not influence timings within \app's execution.

The latency introduced by the System Call Dispatcher to the system calls in \app should also be considered. The measured maximum latency for the system call dispatching was 77 CPU cycles vs. 1 CPU cycle in the baseline FreeRTOS without dispatching. This introduces an overhead of up to 76 cycles for system call dispatching.
There is no overhead for system calls originating from \napp, since they are not dispatched. 

Another potential overhead, although likely infrequent, is from \acro interference exception handling. If \napp tries to access resources in use by \app, an exception is triggered. \acro then logs the interference attempt to the \concept report and releases the resource for \napp to use. Regarding interference exception handling, the maximum time measured from when triggered to when control is returned to the interfering task is 92 CPU cycles.

To evaluate the effects of \acro on RTOS context switches, we measure the time from the completion of the RTOS context switch function to the resumption of the active task. When analyzing the RTOS context switch latency, we focus exclusively on transitions from \napp to \app. We observe the overhead of this transition to be at most 98 CPU cycles. We do not consider the RTOS context switch from \app to \napp because, when it happens, the context initially changes to \napp due to the systick interrupt used by the RTOS scheduler. Consequently, when the PendSV interrupt occurs to switch to a \napp task, the context is already in \napp. Hence, there is no overhead in the measured time.

\begin{table}[H]
\centering
\caption{Overhead Introduced by \acro modules}
\begin{tabular}{l|cc}
\hline
\multicolumn{1}{c|}{}                                 & \multicolumn{2}{c}{\textbf{CPU Cycles}} \\
\multicolumn{1}{c|}{\multirow{-2}{*}{\textbf{Event}}} & \textbf{Max}       & \textbf{Min}       \\ \hline
System Call Dispatching                                   & 77                 & 74                 \\
\rowcolor[HTML]{EFEFEF} 
Initialization Routine                   & 8512             & 8436             \\
Finalization Routine                     & 14332             & 3252             \\
\rowcolor[HTML]{EFEFEF} 
Interference Exception                                & 92                 & 91                 \\
RTOS Context Switch $\napp\rightarrow\app$  & 98&93    \\
\hline
\end{tabular}
\label{table:latency extra}
\end{table}

\subsection{Memory Footprint} 

Table \ref{table:codesize} presents (i) the number of lines of C+Assembly code, measured using "cloc v1.90"; (ii) the corresponding compiled binary size compiled using the "-O0" optimization flag (measured with the STM32CubeIDE build analyzer tool); and (iii) the maximum RAM usage at runtime (measured using the STM32CubeIDE static stack analyzer tool) for each of \acro's modules. \rot implementation in the Secure World (TCB) comprises 569 lines of code. The Initialization and Finalization Routines, excluding the cryptographic libraries, occupy 0.3KBytes and 0.4KBytes of binary size, respectively. Their maximum runtime RAM usage is 272 Bytes and 256 Bytes, respectively. The cryptographic library is striped to include only the code necessary for HMAC and SHA224 usage of built-in cryptographic accelerator hardware in the MCU. This code occupies 2.7KBytes in the binary and allocates 256 Bytes of RAM at runtime. The Interrupt Dispatcher, System Call Dispatcher, and Interference Handler together occupy around 0.9KBytes of PMEM and do not use any RAM, as all memory operations are done at the register level. In the Non-Secure World, the Shadow Task code occupies 0.2KBytes of \napp PMEM and allocates 16 Bytes of RAM at runtime.

\begin{table}[H]
\caption{\acro Memory Footprint}
\begin{tabular}{lccc}
\hline
\rowcolor[HTML]{FFFFFF} 
\multicolumn{1}{c|}{\cellcolor[HTML]{FFFFFF}\textbf{Module}}      & \textbf{\begin{tabular}[c]{@{}c@{}}Binary Size \\ (KBytes)\end{tabular}} & \textbf{\begin{tabular}[c]{@{}c@{}}Lines of \\ Code\end{tabular}} & \textbf{\begin{tabular}[c]{@{}l@{}}Max RAM\\ Usage (Bytes)\end{tabular}} \\ \hline
\rowcolor[HTML]{FFFFFF} 
\multicolumn{4}{c}{\cellcolor[HTML]{FFFFFF}Secure World}                                                                                                                                                                                                                                       \\ \hline
\multicolumn{1}{l|}{System Call Dispatcher}                       & 0.1                                                                      & 28                                                                &           0                                                                  \\
\rowcolor[HTML]{EFEFEF} 
\multicolumn{1}{l|}{\cellcolor[HTML]{EFEFEF}Interrupt Dispatcher} & 0.5                                                                      & 86                                                               &        0                                                                    \\
\multicolumn{1}{l|}{Interference Handler}                         & 0.3                                                                     & 65                                                                &          0                                                                  \\
\rowcolor[HTML]{EFEFEF} 
\multicolumn{1}{l|}{\cellcolor[HTML]{EFEFEF}Finalization Routine} & 0.3                                                                      & 84                                                               &           272                                                                   \\
\multicolumn{1}{l|}{Initialization Routine}                       & 0.3                                                                      & 74                                                               &            256         
\\\rowcolor[HTML]{EFEFEF} 

\multicolumn{1}{l|}{\cellcolor[HTML]{EFEFEF}Cryptography Library}                    & 2.7                                                                      & 232                                                               &            256                                                                 \\ \hline
\rowcolor[HTML]{FFFFFF} 
\multicolumn{4}{c}{\cellcolor[HTML]{FFFFFF}Non-Secure World}                                                                                                                                                                                                                                   \\ \hline
\multicolumn{1}{l|}{Shadow Task}                                 & 0.2                                                                      & 30                                                                &           16                                                                  \\ \hline
\end{tabular}
\label{table:codesize}
\end{table}

 \acro default implementation reserves 2.4Kbytes of RAM to store logs generated during the \concept instances and all metadata contained in the report. This memory was sufficient for all applications considered in our tests
 (see Section~\ref{sec:e2e} for log sizes required by each tested application as a function of interrupt frequencies).

%
%



 

\begin{table*}[]
\centering
\caption{\acro runtime overhead on \app for different number of tasks in \napp with same priviledge}
\begin{tabular}{@{}cc|ccccc|ccc@{}}
\toprule
\multirow{2}{*}{\textbf{\begin{tabular}[c]{@{}c@{}}Number\\  of Tasks\\ \end{tabular}}} & \multirow{2}{*}{\textbf{\begin{tabular}[c]{@{}c@{}}Interrupt \\ Frequency\\ \end{tabular}}} & \multicolumn{5}{c|}{\textbf{Sensor Applications}} & \multicolumn{3}{c}{\textbf{BEEBS Programs}} \\
 &  & \multicolumn{1}{l}{\textbf{Ultrasonic}} & \multicolumn{1}{l}{\textbf{Geiger}} & \multicolumn{1}{l}{\textbf{Syringe}} & \multicolumn{1}{l}{\textbf{Temperature}} & \multicolumn{1}{l|}{\textbf{GPS}} & \multicolumn{1}{l}{\textbf{prime}} & \multicolumn{1}{l}{\textbf{crc32}} & \multicolumn{1}{l}{\textbf{sglib-arraybinsearch}} \\ \midrule
1 & 1 kHz & 15.1\% & 19.8\% & 12.8\% & 11.5\% & 18.7\% & 16.6\% & 10.6\% & 15.2\% \\
2 & 1 kHz & 15.3\% & 20.2\% & 14.4\%  & 11.7\%  & 18.9\% & 17.0\% & 10.8\% & 15.7\% \\
4 & 1 kHz & 15.9\% & 22.7\% & 15.2\%  & 12.9\%  & 19.4\% & 17.5\% & 11.1\% & 16.3\% \\
8 & 1 kHz & 17.5\% & 26.8\% & 17.9\%  & 14.0\%  & 20.8\% & 18.3\% & 11.6\% & 16.7\% \\ \hline

\end{tabular}
\label{tab: overhead number of tasks}
\end{table*}

\begin{table*}[]
\centering
\caption{\acro runtime overhead on \app task for different interrupt frequency in \napp}
\begin{tabular}{@{}cc|ccccc|ccc@{}}
\toprule
\multirow{2}{*}{\textbf{\begin{tabular}[c]{@{}c@{}}Number\\  of Tasks\\ \end{tabular}}} & \multirow{2}{*}{\textbf{\begin{tabular}[c]{@{}c@{}}Interrupt \\ Frequency\\ \end{tabular}}} & \multicolumn{5}{c|}{\textbf{Sensor Applications}} & \multicolumn{3}{c}{\textbf{BEEBS Applications}} \\
 &  & \multicolumn{1}{l}{\textbf{Ultrasonic}} & \multicolumn{1}{l}{\textbf{Geiger}} & \multicolumn{1}{l}{\textbf{Syringe}} & \multicolumn{1}{l}{\textbf{Temperature}} & \multicolumn{1}{l|}{\textbf{GPS}} & \multicolumn{1}{l}{\textbf{prime}} & \multicolumn{1}{l}{\textbf{crc32}} & \multicolumn{1}{l}{\textbf{sglib-arraybinsearch}} \\ \midrule
1 & 2 kHz & 15.2\% & 19.8\% & 12.9\% & 11.5\% & 18.8\% & 16.6\% & 10.6\% & 15.2\% \\
1 & 4 kHz & 15.4\% & 19.9\% & 13.0\% & 11.5\% & 18.9\% & 16.7\% & 10.6\% & 15.3\% \\
1 & 6 kHz & 15.7\% & 20.2\% & 13.3\% & 11.6\% & 19.0\% & 16.7\% & 10.6\% & 15.5\% \\
1 & 8 kHz & 16.0\% & 20.5\% & 13.7\% & 11.8\% & 19.2\% & 16.9\% & 10.7\% & 15.9\% \\ \hline
\end{tabular}
\label{table: interrupt freq variating}
\end{table*}

\begin{table*}[]
\centering
\caption{\acro runtime overhead on \napp task for different interrupt frequency in \app}
\begin{tabular}{@{}cc|ccccc|ccc@{}}
\toprule
\multirow{2}{*}{\textbf{\begin{tabular}[c]{@{}c@{}}Number\\  of Tasks\\ \end{tabular}}} & \multirow{2}{*}{\textbf{\begin{tabular}[c]{@{}c@{}}Interrupt \\ Frequency\\ \end{tabular}}} & \multicolumn{5}{c|}{\textbf{Sensor Applications}} & \multicolumn{3}{c}{\textbf{BEEBS Applications}} \\
 &  & \multicolumn{1}{l}{\textbf{Ultrasonic}} & \multicolumn{1}{l}{\textbf{Geiger}} & \multicolumn{1}{l}{\textbf{Syringe}} & \multicolumn{1}{l}{\textbf{Temperature}} & \multicolumn{1}{l|}{\textbf{GPS}} & \multicolumn{1}{l}{\textbf{prime}} & \multicolumn{1}{l}{\textbf{crc32}} & \multicolumn{1}{l}{\textbf{sglib-arraybinsearch}} \\ \midrule
1 & 2 kHz & 13.5\% & 15.2\% & 10.8\% & 10.5\% & 14.2\% & 13.3\% & 10.7\% & 12.7\% \\
1 & 4 kHz &13.5\% & 15.3\% & 10.8\% & 10.5\% &  14.3\% & 13.4\% & 10.7\% & 12.8\% \\
1 & 6 kHz & 13.7\% & 15.5\% & 11.0\% & 10.6\% & 14.6\% & 13.4\% & 10.8\% & 13.1\% \\
1 & 8 kHz & 13.7\% & 16.1\% & 11.3\% & 10.7\% & 15.2\% & 13.6\% & 10.8\% & 13.5\% \\ \hline

\end{tabular}
\label{table: interuupt freq from app}
\end{table*}

\input{tables/report_size.tex}

\subsection{End-To-End Run-Time Evaluation}\label{sec:e2e}

We assess the performance of \acro on real embedded applications through a series of experiments involving a selection of open-source MCU applications and MCU programs from the BEEBS benchmark \cite{beebs}. Initially, we focus on a case where only one task is being scheduled by the RTOS using a systick interrupt configured to trigger at 1 kHz, running without interference from other tasks. In this setup, we aim to assess the execution delay experienced by the task both with and without \acro. We evaluate eight different algorithms under conditions where no additional interrupts are set up for the task.

The results of the experiment is illustrated in the first row of Table  \ref{tab: overhead number of tasks}. When \acro is activated, the task experiences a runtime increase that varies from 10.6\% to 19.8\%, reflecting the overhead introduced by the additional security processes. This increase is relatively modest across all eight algorithms tested, suggesting that \acro efficiently integrates security without significantly compromising the system's performance.

Next, we expand our evaluation to include scenarios where the RTOS schedules multiple tasks simultaneously.
In this configuration, all tasks are assigned the same priority, ensuring the RTOS allocates the same time budget to each. We vary the number of tasks within \napp from 1 to 8, focusing on assessing the overhead generated by the task operating within \app, while maintaining the \napp interrupt frequency fixed at 1 kHz. By analyzing the results in Table \ref{tab: overhead number of tasks}, we observe that overhead increases across all applications, indicating an approximately linear increase in overhead relative to the number of tasks. 

Next, we evaluate how \acro performs when a task running in \app is subjected to interrupts at frequencies ranging from 1kHz to 8kHz in \napp. This test aims to assess the scalability of \acro under varying levels of interrupt-driven stress. The results of this experiment are presented in Table \ref{table: interrupt freq variating}. We observe a linear increase of approximately 0.1\% overhead increase for each 1kHz increment in the interrupt frequency.

We also evaluate the overhead of a task in \napp when there is an active \concept instance on \prv. We consider a single task running in \napp and another in \app. We vary the interrupt frequency within \app from 2kHz to 8kHz, while \napp maintains a fixed systick frequency of 1kHz. The results of these experiments are presented in Table \ref{table: interuupt freq from app}. By comparing Table \ref{table: interrupt freq variating} with Table \ref{table: interuupt freq from app}, we observe that for the same interrupt frequency, the runtime overhead for tasks in \napp is lower than the overhead generated by tasks in \app. However, in both cases, we note that the increase in overhead is proportional to the increase in interrupt frequency. Both present similar growth rates.

Finally, Table \ref{table: log size} presents \acro report sizes across different applications, with varying \napp interrupt frequency. 
Higher interrupt frequencies lead to more context switches \app $\leftrightarrow$ \napp, resulting in a higher number of \napp timings logged and thus a larger log size.
Log sizes across tested applications ranged from 84 to 244 bytes.

\section{Related Work}\label{sec:rw}

\textbf{Remote Attestation (\RA):}
\RA architectures are generally classified in three types: software-based (or keyless), hardware-based, and hybrid.
Software-Based \RA Architectures ~\cite{KeJa03, SPD+04, SLS+05, SLP08, pistis, simple, surminski2021realswatt, scraps} do not rely on any specialized hardware. The main advantage of software-based \RA is its flexibility and ease of implementation since it does not require additional hardware. However, this model places significant demands on the security of the software itself. It operates under strict assumptions about the capabilities of \adv capabilities and the flawless implementation of the \RA mechanism. This can be a limiting factor as it may not provide robust security against sophisticated hardware-level attacks.
Hardware-Based RA Architectures~\cite{PFM+04, KKW+12, SWP08, sacha, Sancus17} incorporate dedicated hardware to support security features, such as Trusted Platform Modules (TPMs)~\cite{tpm} or uses specific CPU instruction sets like Intel SGX~\cite{sgx-explained}. The use of hardware provides a stronger security foundation, as the \RA processes are less vulnerable to software attacks and can leverage the physical security properties of the hardware. The primary drawback is the increased cost and complexity of integrating specialized hardware components, which might not be feasible for all devices or contexts.
Hybrid RA Architectures~\cite{vrased, smart, tytan, trustlite, hydra, brasser2016remote} aim to bridge the gap between software and hardware solutions. This type of design seeks to combine the lower cost and flexibility of software-based RA with the robust security features of hardware-based RA. This is often achieved through hardware/software co-designs that smartly integrate both approaches to fit specific use cases. For example, hybrid \RA might use minimal hardware to bolster a primarily software-based \RA process, thus providing enhanced security without the full cost of hardware-based systems.

\textbf{Proof of Execution (\pox):} 
On high-end devices (application computers and servers), Flicker \cite{mccune2008flicker} was the first to utilize hardware support for a dynamic root of trust for measurement in TPMs (along with AMD’s SVM support in commodity AMD processors) to prove execution of code in isolation. A Flicker session sets an isolated and uninterruptable environment (protected mode) for the execution to be proven in order to prevent interference from other software on the same machine. Also in high-end devices, Sanctum \cite{costan2016sanctum} employs a similar approach by instrumenting Intel SGX’s enclave code to convey information about its execution to a remote party. APEX~\cite{apex} introduces the notion of \pox on low-end single-core devices running bare-metal software. It implements a verified hardware monitor to check for violations to the atomic execution of the application being attested, disallowing interrupts during the execution being proven. ASAP~\cite{caulfield2022asap} extends APEX to allow the proven task to implement its own interrupts but still disallows interrupts external to the task being proven.

\textbf{Control Flow Attestation (\CFA):}  
\CFA generates remotely verifiable evidence for the control flow path taken by an application running on \prv~\cite{ammar2024sok}. This is achieved by instrumenting the application code \cite{oat,iscflat,cflat,scarr,recfa,wang2023ari} with instructions to track the control flow path during a \pox. This process generates a measurement that verifies the execution of each node in the control flow graph. Alternatively, hardware-based techniques \cite{tinycfa,lofat,dessouky2018litehax} leverage custom hardware to track the control flow path.
In the context of real-time systems, ISC-FLAT \cite{iscflat} safeguards CFA logged control flow transfers by allowing interrupts while blocking critical resources during the interruptions. However, this approach conflicts with RT-PoX because blocked resources cannot be managed or accessed by the RTOS, impeding co-existence.
ARI \cite{wang2023ari} offers a CFA mechanism that includes timing information in CFA reports. However, it does not address coexistence between an unmodified (Non-Secure World-resident) RTOS and \pox.


\textbf{Control Flow Integrity (\CFI):} 
In contrast to \CFA, \CFI techniques implement measures to detect control flow path violations in place, locally on \prv. Different from \CFA (and \pox), \CFI is not concerned with the generation of remotely verifiable execution evidence. Static \CFI uses static analysis at compilation time to create a control flow graph (CFG) and insert runtime checks to ensure that control transfers match the CFG~\cite{tice2014enforcing,van2015practical,walls2019control}. For instance, Microsoft's Control Flow Guard \cite{MicrosoftCFG} adds runtime checks to validate indirect function calls for forward-edge protection.
Dynamic CFI \cite{nyman2017cfi,christoulakis2016hcfi}, monitors the control flow at runtime beyond static CFG, often using hardware support~\cite{sherloc}. Return Address Validation and Shadow Call Stacks protect return addresses on the program's stack from tampering \cite{muRAI,burow2019sok, zhou2020silhouette}.


\textbf{Security in Real-Time Systems.} Various prior works enhance the security of real-time systems in aspects other than proving code or execution integrity. Scheduler-based methods leverage the scheduler to monitor task execution and impose security restrictions \cite{hao2019integrating,hasan2016exploring,hamad2018prediction} or to decrease system predictability, thereby mitigating side-channel attacks \cite{pellizzoni2015generalized,yoon2016schedule,kruger2018vulnerability}.
Some techniques use time reservation from conservative systems to enforce data flow integrity \cite{wangopportunistic} or implement runtime monitors \cite{10.1145/3578359.3593038}. Task memory isolation to reduce the attack surface is also studied in \cite{kim2018securing}.

\textbf{Availability in Real-time Systems.}
Recent efforts have focused on providing availability guarantees in real-time systems. \cite{aion} and \cite{garota} propose architectural mechanisms to fuse interrupts (including time-based) with isolated execution of trusted software despite full software compromise of MCUs.
\cite{proactive3} proposes the concept of trusted scheduling, i.e., leveraging trusted computing to ensure real-time guarantees. 
More recently, \cite{wang2022rt} leverages hardware primitives on commodity platforms to ensure the availability of system resources (e.g., CPU and I/O) for real-time tasks, along with trusted scheduling of these tasks. 
\concept goal differs from previous work as it aims to enable the coexistence of off-the-shelf RTOS functions with PoX rather than enforcing trusted scheduling or guaranteed availability (see Appendix C for additional discussion). 

\section{Conclusion}

We formulated the \concept concept to address the limitations of classic \pox definitions and make them compatible with real-time systems. To realize \concept, we developed \acro as the first \pox architecture compatible with an off-the-shelf RTOS (FreeRTOS). This enables both real-time availability offered by the RTOS and \pox verifiable integrity guarantees to co-exist securely on off-the-shelf MCUs. We also evaluate \acro's security and performance overheads and make its prototype publicly available~\cite{repo}.
\concept formulation and its instantiation in \acro open several avenues for future work. Among them, we highlight the following.
 
\textbf{Mitigating vulnerabilities within \app:}
\concept is designed to protect an application in \app from external adversaries in \napp. However, when \app is itself potentially vulnerable (e.g., \app has a memory safety bug) \adv can manipulate its control flow path within the \pox. To effectively enforce or remotely verify the control flow of \app, \acro can be used alongside other techniques such as \CFI and \CFA. See Section~\ref{sec:rw} for an overview of \CFI and \CFA.



\textbf{\app Confidentiality:} \pox focuses on providing integrity while \concept augments \pox not to violate the availability needs of RTOS-based environments. Our treatment does not consider systems in which confidentiality of data is also a pressing concern. For a discussion of the data confidentiality aspect of potentially compromised embedded/sensing systems see~\cite{nunes2022privacy}. Incorporating confidentiality into \concept is an interesting avenue for future work.


\label{sec:report scalability}
\textbf{Report Scalability}. \acro logs timing events to detect interferences, as detailed in Section~\ref{sec:interference}. 
This implies that extended executions could generate a large log size. To mitigate this, optimizations can be implemented, e.g., sending a partial report to \vrf when \prv's memory fraction budgeted for storing logs is full. Another alternative is \vrf  to implicitly specify acceptable pause times for \app and have \acro check if \vrf policy is met during the \concept, reporting a yes/no result to \vrf.

\section*{Acknowledgements}

We sincerely thank the paper's shepherd and anonymous reviewers for their feedback and guidance.
This work was partly supported by the National Science Foundation (SaTC award \#2245531).
PSU author was partly supported by the ASEAN IVO (\url{www.nict.go.jp/en/asean_
ivo/}) project, Artificial Intelligence Powered Comprehensive Cyber-Security for Smart Healthcare Systems (AIPOSH), funded by NICT (\url{www.nict.go.jp/en/}).

%% file: tables/report_size.tex
\begin{table*}[ht]
\centering
\caption{\acro Log Size (bytes) for varying interrupt frequencies in \napp}
\begin{tabular}{@{}cc|ccccc|ccc@{}}
\toprule
\multirow{2}{*}{\textbf{\begin{tabular}[c]{@{}c@{}}Number\\  of Tasks\\ \end{tabular}}} & \multirow{2}{*}{\textbf{\begin{tabular}[c]{@{}c@{}}Interrupt \\ Frequency\\ \end{tabular}}} & \multicolumn{5}{c|}{\textbf{Sensor Applications}} & \multicolumn{3}{c}{\textbf{BEEBS Applications}} \\
 &  & \multicolumn{1}{l}{\textbf{Ultrasonic}} & \multicolumn{1}{l}{\textbf{Geiger}} & \multicolumn{1}{l}{\textbf{Syringe}} & \multicolumn{1}{l}{\textbf{Temperature}} & \multicolumn{1}{l|}{\textbf{GPS}} & \multicolumn{1}{l}{\textbf{prime}} & \multicolumn{1}{l}{\textbf{crc32}} & \multicolumn{1}{l}{\textbf{sglib-arraybinsearch}} \\ \midrule
1 & 2 kHz & 104 B  & 84 B  & 84 B  & 84 B  &  84 B & 124 B & 104 B  & 104 B \\
1 & 4 kHz & 124 B & 104 B  &104 B  & 84 B  & 104 B & 164 B & 124 B &  124 B \\
1 & 6 kHz & 144 B & 144 B & 144 B & 104 B &  144 B & 204 B & 144 B &  144 B \\
1 & 8 kHz & 184 B & 184 B & 184 B & 144 B &  184 B & 244 B & 184 B &  184 B \\ \hline
\end{tabular}
\label{table: log size}
\end{table*}

%% file: pages/15_Appendix.tex





\appendices 

\section{Shadow Task Details }\label{apdx:shadow}

\begin{figure}[!h]
    \centering
    \includegraphics[width=1\linewidth]{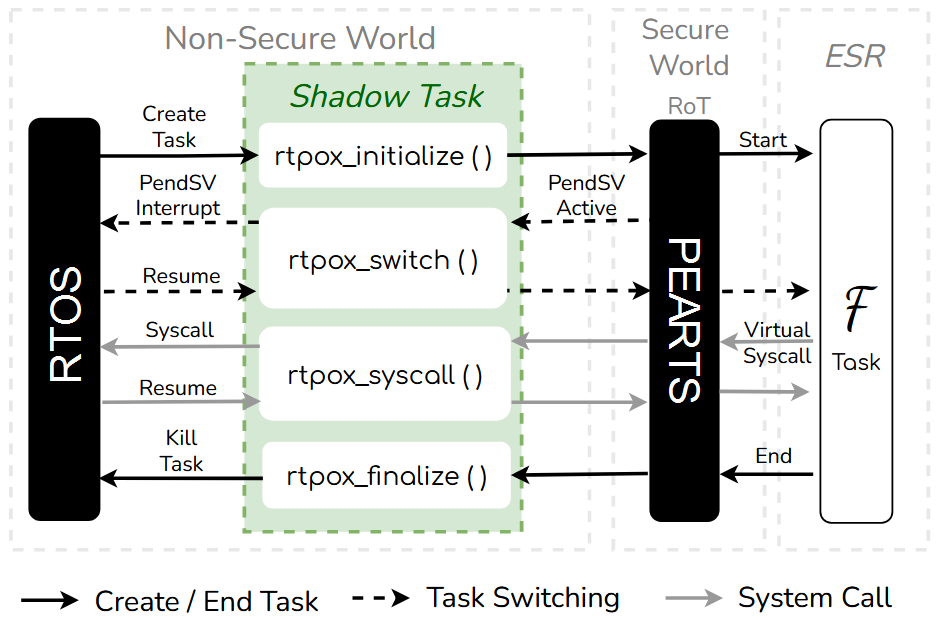}
    \caption{Shadow Task interfacing the RTOS and \acro.}
    \label{fig:shadow task}
\end{figure}

A shadow task is created by the RTOS in the Non-Secure World to serve as a user-space handle used by the RTOS to interact with \app, since \app is isolated from \napp. It has three main functionalities: (1) starting and terminating an \concept process, (2) handling system calls, and (3) intermediating task switching by the RTOS.
Figure~\ref{fig:shadow task} illustrates the interplay between the shadow task and other components in \acro.

\textbf{Starting/Terminating \concept}. When \vrf requests \concept of \app, the RTOS creates the shadow task in the Non-Secure World, whose initial behavior is to trigger \acro to start the \acro Initialization Routine (see Section~\ref{sec:init}).  When the \concept process is finished, \acro sends the report to the shadow task. The shadow task then forwards the report either directly to \vrf or to another task responsible for the report transmission. After that, the shadow task asks the RTOS to terminate the task of \app.

\textbf{Handling System Calls.} Since \napp is untrusted, \acro must prohibit the use of system calls that return data from \napp to \app, as this could affect \app's behavior (recall Property 3). To enforce this, during the offline phase, all system calls made by \app are replaced with virtual instructions that invoke the \acro system call dispatcher. 
When \app triggers a system call, the \acro system call dispatcher (1) copies the arguments to the Non-Secure World context that will later be used by the shadow task; (2) logs the event in the \concept report (including the system call type and arguments); and (3) resumes the shadow task in the Non-Secure World. When the CPU changes to Non-Secure mode, the registers are loaded with the system call context and the shadow task calls the corresponding system call. When system calls return `void', the Shadow Task immediately triggers \rot to resume \app. 
If the system call returns a status (success or failure), the Shadow Task relays this information to \acro dispatcher, which logs it in the report before sending it to \app.
For synchronization, tasks using mutexes and semaphores must hold the corresponding handles.
In this case, the shadow task holds the handle while the \acro system call dispatcher creates a token representing it and gives this token to \app. When \app needs to use the handle, it presents the token to the \acro system call dispatcher, which subsequently signals the shadow task to use the corresponding object on \app's behalf.

\textbf{Task Switching.} The shadow task acts as an intermediary for task switching by the RTOS. Suppose the CPU is executing \app inside the ESR during \concept. When a systick interrupt occurs, \acro dispatches it, switches the context to the Non-secure World and directs execution to the systick interrupt handler of the RTOS. Within the handler execution, the RTOS checks if a task switch is needed. If so, it sets the PendSV flag and then exits.
After the systick handler ends and returns the control to \acro, \acro checks if the PendSV flag is active. If it is not, \acro resumes \app. When the flag is active, it means the RTOS wants to switch tasks. In this case, \acro transfers control to the Shadow Task, which is immediately interrupted by the PendSV interrupt. 
The PendSV interrupt handler then switches the context to another task.
When the RTOS decides to switch back to \app, it first restores the Shadow Task’s context and resumes its execution. The Shadow Task then invokes \acro to resume \app.


\section{System Calls Available to \app }\label{apdx:syscall}


During an \concept process, to ensure Property 2, \acro does not expose all FreeRTOS system calls to \app. Specifically, any system call that fetches data (e.g., peripheral inputs) to \app must be restricted because a malicious RTOS could provide tampered data, compromising \app's context. To cope with this, \app must implement its own data fetching functions (which is feasible and simple in bare-metal MCUs). Other than data fetching, \acro supports the use of FreeRTOS system calls in two categories.

The first category is \textbf{control-yield system calls}, which allows \app to yield control of its execution. An example is ``vTaskDelay()'', where the task requests the RTOS to pause its execution for a certain amount of time. While a malicious RTOS could misbehave and delay for an incorrect amount of time, this behavior would be measured and subsequently detected by \vrf in \acro. This is because \acro measures time independently when \app's execution is paused.

The second category includes \textbf{synchronization system calls}, used when \app needs to synchronize with another task in \napp to access shared resources such as peripherals. These calls are allowed because synchronization controls resource access but does not involve sharing data and invalid access conflicts would be detected by \acro. For example, if the RTOS falsely signals that a resource is free when it's not, and \napp modifies it while \app is using it, an interference exception would occur and be detected by \vrf. Conversely, if the RTOS falsely claims a resource is never available, \vrf would detect this through timing measurements that reveal the RTOS's blocking behavior.

A third category of system calls (although not currently implemented in \acro prototype) is possible: \textbf{data-transmission system calls}. These system calls could be allowed when originating from \app since sending data to a task in \napp does not violate Property 2. In this case, the associated Shadow Task would manage the transfer, where the system call dispatcher copies the data to the shadow task. Alternatively, a shared memory region could be used to send the data. As mentioned above, \app should not blindly receive and use data fetched/generated by \napp, so data-receiving system calls are not allowed.

\label{subsec:more datils on Wrapper Taks}
\label{subsection:interface}

\section{ 
\concept vs. Trusted Scheduling 
}\label{sec:appendix:availability}

Property 5 in Section~\ref{subsec:requirements} states ``RTOS availability" as the ability for \pox to function alongside the RTOS without disrupting its underlying duties or giving up on the system's timing requirements outside the context of \app.
\acro realizes this by never blocking essential operations and resources, not interfering with RTOS task scheduling and resource management mechanisms, and keeping \concept-related overheads small.

The goal above is not to be confused with prior work on mechanisms that support/leverage trusted scheduling~\cite{proactive3,garota,aion,wang2022rt}. The latter aims to guarantee real-time availability despite compromised software states.
\concept, on the other hand, enables remotely verifiable execution of a specific task (\app), defined and requested by \vrf. The evidence contained in an \concept report reliably informs \vrf of interruptions (and their durations) during the execution of \app, as well as any interference in \app's context through its provable execution. Different from the related work above, \acro does not involve bringing functions that would normally be performed by the Non-Secure World RTOS (e.g., scheduling of Non-Secure World tasks) into the Secure World to guarantee their occurrence. Instead, it allows the RTOS to remain in the Non-Secure World while enabling remote verifiability of execution integrity and associated timing for specific tasks upon \vrf's request.


\ignore{
The availability we aim to provide when refers to RTOS availability is the ability of our system to function alongside the RTOS without disrupting its operations or compromising its timing requirements. Our system is designed to support the RTOS by maintaining system integrity and ensuring real-time capabilities, such as predictable scheduling and task execution, are not hindered. In this context, availability means that our system does not introduce significant delays, block essential operations, or interfere with the RTOS's task scheduling and resource management mechanisms.

It is important to clarify that, while important, our goal is not to guarantee the availability of individual tasks or their execution outcomes within the RTOS.
We leave the responsibility for managing task availability, scheduling, and execution to the RTOS itself. The RTOS handles task prioritization, resource allocation, and synchronization between tasks based on its own scheduling algorithms and design goals.
}

\label{subsec: available syscalls}
\begin{table}[H]
\centering
\caption{List of FreeRTOS system calls allowed by \acro}
\resizebox{\columnwidth}{!}{%
\begin{tabular}{@{}ll@{}}
\toprule
\textbf{FreeRTOS Services} &
\textbf{System Calls Available to \app} \\ \midrule
Task Control &
\begin{tabular}[c]{@{}l@{}}vTaskDelay()\\ vTaskDelayUntil()\\ xTaskDelayUntil()\\ vTaskPrioritySet()\\ vTaskSuspend()\\ vTaskResume()\\ xTaskAbortDelay()\end{tabular} \\ \midrule
Direct To Task Notifications &
\begin{tabular}[c]{@{}l@{}}xTaskNotifyGive()\\ xTaskNotify()\\ xTaskNotifyAndQuery()\\xTaskNotifyWait()\\ xTaskNotifyStateClear()\end{tabular} \\ \midrule
Semaphore / Mutexes &
\begin{tabular}[c]{@{}l@{}}xSemaphoreCreateBinary()\\ xSemaphoreCreateBinaryStatic()\\ vSemaphoreCreateBinary()\\ xSemaphoreCreateCounting()\\ xSemaphoreCreateCountingStatic()\\ xSemaphoreCreateMutex()\\ xSemaphoreCreateMutexStatic()\\ xSemaphoreCreateRecursiveMutex ()\\ xSemaphoreCreateRecursiveMutexStatic()\\ vSemaphoreDelete()\\ xSemaphoreTake()\\ xSemaphoreTakeRecursive()\\ xSemaphoreGive()\\ xSemaphoreGiveRecursive()\end{tabular} \\ \midrule
\end{tabular}%
}
\end{table}